\documentclass[12pt,a4paper]{article}
\usepackage{a4wide}
\usepackage{amsmath}
\usepackage{amsfonts}
\usepackage{cite}
\usepackage{graphicx}
\begin{document}

\title{On the non-$k$-separability of Dicke class of states and $N$-qudit W states}
\author{{N. Ananth and  M. Senthilvelan }\\
{\small Centre for Nonlinear Dynamics, School of Physics, Bharathidasan University,} \\{\small Tiruchirappalli - 620024, Tamilnadu, India} }
\date{}
\maketitle

\begin{abstract}
In this paper, we present the separability criteria to identify non-$k$-separability and 
genuine multipartite entanglement 
in mixed multipartite states using elements of density matrices. 
Our criteria can detect the non-$k$-separability of Dicke class of states, anti W states and  mixtures thereof and 
higher dimensional W class of states. 
We then investigate the performance of our criteria by considering $N$-qubit Dicke states with arbitrary excitations 
added with white noise and mixture of $N$-qudit W state with white noise. We also study the 
robustness of our criteria against white noise. 
Further, we demonstrate that our criteria are experimentally implementable by means of 
local observables such as Pauli matrices and generalized 
Gell-Mann matrices.
\end{abstract}

\section{Introduction}
\label{intro}
One of the important areas of research in quantum information is to identify  
multipartite entanglement in the arbitrary multipartite states \cite{horo2009,guhne2009}. 
To detect multipartite entanglement two approaches have been widely used, namely   
(i) $k$-separability criteria and (ii) $n$-party entanglement conditions \cite{guhne2005}.  
The multipartite entangled states that are invariant under permutations are useful 
for the quantum information processing \cite{novo2013}. 
The Greenberger-Horne-Zeilinger (GHZ) states, Dicke states (including $N$-qubit W states) and higher dimensional W states are some notable examples in the permutationally invariant states. 
In this paper, we formulate the non-$k$-separability criteria for the $N$-qubit Dicke states with 
arbitrary excitations and $N$-qudit W states. 
  
\subsection{Dicke states}
Dicke states which were first observed from spontaneous emission of light by a cloud of atoms \cite{dick1954} 
are the family of multiqubit entangled states 
which are symmetric with respect to the subsystem permutations. 
An $N$-qubit Dicke state with $m$ excitations is defined as \cite{toth2007,berg2013} 
\begin{align}
|D_m^N \rangle = \frac{1}{\sqrt{({N\atop m})}} \sum_j P_j \Big\{|1\rangle^{\otimes m}\otimes |0\rangle^{\otimes(N-m)}\Big\}, 
\end{align}
where $\sum_j P_j \{.\}$ denotes the sum over all possible permutations. For example, 
a $4$-qubit Dicke state with $3$ excitations is denoted by 
$|D_3^4\rangle = \frac{1}{\sqrt{4}}\Big(|0111\rangle + |1011\rangle$ $+ |1101\rangle + |1110\rangle \Big)$.  

Dicke states have been generated and studied experimentally with systems like photons and ions 
\cite{haff2005,kies2007,prev2009,wiec2009,lini2008,toyo2011}. 
Since Dicke states are most robust against decoherence \cite{guhn2008}, 
they can be used effectively for certain tasks such as open-destination teleportation \cite{kies2007}, 
quantum telecloning \cite{mura1999} and quantum secret sharing \cite{hill1999}. 
Recently, various tools have been developed to detect genuine multipartite entanglement in Dicke states and 
in the vicinity of Dicke class of states. 
To name a few, we cite    
(i) fidelity-based entanglement witness operators \cite{toth2007},  
(ii) set of inequalities which involve permutation operators \cite{hube2011},  
(iii) the criteria which based on simple measurements of collective spin operators \cite{duan2011}, 
(iv) entanglement witnesses using the method of PPT mixtures \cite{berg2013} and  
(v) the criterion which based on the measurement of global spin \cite{luck2014}. 
The criterion which isolates the non-$k$-separability in Dicke class of states with 
single excitation was derived in \cite{gao2014,anan2015}. 

\subsection{Higher dimensional W states}
The $N$-qudit W state has several generalizations \cite{gabr2010,kim2008}. 
One such generalization, namely a $3$-qutrit W state ($1/\sqrt{6}\left(|012\rangle + |021\rangle + |102\rangle + |120\rangle + |201\rangle +|210\rangle \right)$),  
was considered in \cite{gabr2010}. 
The $3$-qutrit W state has been generalized to $N$-qudit W state $(d=N)$ as 
\begin{align}
|W_N^d\rangle = \frac{1}{\sqrt{N!}} \sum_i P_i \Big\{ | 012\ldots (d-2)(d-1)\rangle\Big\}, \label{nqdw}
\end{align}
where $\sum_i P_i \{.\}$ denotes the sum over all possible permutations. 
In this generalization each system in an $N$ number of system can have $0$ to $N-1$ excitations with a restriction that two subsystems cannnot be in the same excitation at a moment. 
For example, a 4-dimensional 4-partite W state is by definition $|W_4^4\rangle = \frac{1}{\sqrt{24}}\big(|0123\rangle + |0132\rangle + |0213\rangle + |0231\rangle + |0312\rangle 
+ |0321\rangle + |1023\rangle + |1032\rangle + |1203\rangle + |1230\rangle + |1302\rangle + |1320\rangle + |2013\rangle + |2031\rangle + |2103\rangle + |2130\rangle 
+ |2301\rangle + |2310\rangle + |3012\rangle + |3021\rangle + |3102\rangle + |3120\rangle + |3201\rangle + |3210\rangle\big)$.  
W class of entangled states are found applications in quantum teleportation \cite{agra2006}, super dense coding, splitting quantum information \cite{zheng2006} and 
solving quantum leader election problem in networks \cite{hond2006}. 
Several conditions were proposed to detect genuine multipartite entanglement and 
nonseparability in W class of states \cite{anan2015,gabr2010,gao2013,guhne2010,gao2011,hube2010}. 
For example, Huber et al. have proposed a general framework to identify genuinely multipartite entangled 
mixed quantum states in arbitrary dimensional systems \cite{hube2010}. 
Gabriel et al. have developed a necessary criterion for $k$-separable mixed multipartite states \cite{gabr2010}. 
The $k$-nonseparability criteria for the W states and anti W states were developed in \cite{gao2013}. 
Very recently we have formulated the non-$k$-separability criterion for 
a generalized $N$-qudit W state \cite{anan2015,kim2008}. 

Even though attempts have been made to isolate genuine multipartite entanglement and nonseparability  
of these two classes of states, the necessary and/or sufficient criteria to identify the 
non-$k$-separability in Dicke class of states with $m$ excitations and 
for the higher dimensional W states are yet to be formulated. 
Motivated by this observation, in this paper, we propose a set of conditions which identify 
the non-$k$-separability in these two  
mixed multipartite states using elements of density matrices. 
With the help of our criteria one can detect the non-$k$-separability in Dicke class of states, anti W states 
and mixtures thereof and $N$-qudit W class of states. 
We also illustrate the method of identifying genuine multipartite entanglement and nonseparability of 
$N$ qubit Dicke state added with white noise and mixture of $N$-qudit W state with white noise. 
Further we analyze the white noise tolerance of our criteria. 
In addition to the above, we demonstrate that the criteria presented in this paper can be experimentally implementable with the help of local expectation values of Pauli operators and generalized Gell-Mann matrices.   

The structure of this paper is as follows: 
In Section \ref{sec2}, we recall the definition of $k$-separability and give 
an expression to count the number of possible partitions that can exist in the $k$-separable $N$-partite states. 
In Section \ref{sec3}, we present the criteria which identify the non-$k$-separability 
of Dicke class of states and $N$-qudit W class of states. 
The performance of our criteria in identifying non-$k$-separability in the arbitrary 
Dicke state with $m$ excitations added with white noise and mixture of $N$-qudit W state with white noise and  
their robustness against white noise are discussed in Section \ref{sec4}.   
In Section \ref{sec6}, we discuss the experimental feasibility of our criteria in terms of local observables. 
Finally, in Section \ref{sec7}, we summarize our conclusions. 
The proof of one of our criteria is given in Appendix. 
\section{Counts on partitions of $k$-separable $N$-partite states}
\label{sec2}
To begin, we recall the definition of $k$-separability. 
An $N$-partite pure quantum state 
$|\psi_{k-\textrm{sep}}\rangle$ is called $k$-separable 
$(k=2,3,\ldots,N)$ if and only if it can be written as a product of $k$ substates, 
$|\psi_{k-\textrm{sep}}\rangle = |\psi_1\rangle \otimes |\psi_2\rangle \otimes \ldots \otimes|\psi_k\rangle$,    
where $|\psi_i\rangle$, $i=1,2,\ldots,k$, represents the state of a single subsystem or a group of subsystems \cite{gabr2010}.  
A mixed state $\rho_{k-\textrm{sep}}$ is called $k$-separable, if it can be decomposed into pure $k$-separable states, that is 
\begin{align}
\label{k2} \rho_{k-\textrm{sep}} = \sum_i p_i~ \rho_{k-\textrm{sep}}^i,  
\end{align}
where $p_i > 0$ and $\sum_i p_i = 1$. 
An $N$-partite state is non-$k$-separable if it is not $k$-separable \cite{gabr2010}. 
The $\rho_{k-\textrm{sep}}^i$'s are need not be in the same partition - they may be under different partitions. 

In the following, we derive an expression that count the number of possible partitions 
that can exist in the $k$-separable $N$-partite states. 
Let $m_1,m_2,\ldots,m_k$ be the number of subsystem(s) in each compartment of any partition belongs to the 
$k$-separable $N$-partite state  with $\sum_{i=1}^k m_i = N$, $1\leq m_i \leq N$. 
Before going for a general case let us consider a specific example and count 
the number of different partitions it can admit. 
Let us consider a $3$-separable $6$-partite state and count the total number of partitions in it.  
The possibilities in the $3$-separable case are  
(i) $m_1=3$, $m_2=2$, $m_3=1$, (ii) $m_1=1$, $m_2=1$, $m_3=4$ and (iii) $m_1=2$, $m_2=2$, $m_3=2$. 
In the first case ($m_1=3$, $m_2=2$, $m_3=1$), partitions like $ABC|DE|F$, the number of partitions can be $\frac{N!}{m_1! m_2! m_3!}$ , 
that is $\frac{6!}{3! 2! 1!}=60$.     
Suppose the same $6$-partite state is $3$ separable like $A|B|CDEF$ (second case) then 
the number of possible partitions turn out to be $\frac{N!}{2! m_1! m_2! m_3!}$. 
An extra factor $2!$ has been introduced in the denominator since  $m_1=m_2$. 
Since we have $m_1=1$, $m_2=1$ and $m_3=4$, in the present case, the number of possible partitions become $\frac{6!}{2! 1! 1! 4!}=15$. 
The third case, that is $m_1=2$, $m_2=2$ and $m_3=2$, partitions like $AB|CD|EF$, gives us $\frac{N!}{3! m_1! m_2! m_3!} = 15$ number of partitions.   
Considering all three possible partitions we find that the total number of partitions of   
the $3$-separable $6$-partite states should be $90$. 
In a similar manner, to obtain the total number of partitions of the $k$-separable $N$-partite state  
we should consider all possibilities with $m_a + m_b + \ldots + m_k = N$ and their coincidences, 
that is $m_a = m_b = m_c$, $m_a = m_b$ \& $m_c = m_k$ and so on, and count all of them.  
We find that the total number of partitions of the $k$-separable $N$-partite state 
$(N_{\textrm{part}}^k)$ should be  
\begin{align}
N_{\textrm{part}}^k =& \sum_{m_a\neq m_b \neq\ldots\neq m_k} \frac{N!}{m_a! m_b!\ldots m_k!}+\sum_{m_a = m_b} \frac{N!}{2! m_a! m_b!\ldots m_k!} \nonumber \\ 
&+\sum_{m_a = m_b = m_c} \frac{N!}{3! m_a! m_b!\ldots m_k!} + \sum_{m_a = m_b, \atop m_c=m_d=m_e} \frac{N!}{2! 3! m_a! m_b!\ldots m_k!} + \ldots  \nonumber\\ 
&+ \sum_{m_a = m_b = \ldots = m_k} \frac{N!}{k! m_a! m_b!\ldots m_k!},
\end{align}
where $\{a,b,c,\ldots,k\} = \{1,2,\ldots,k\}$ and $\{m_a,m_b,\ldots,m_k\} = \{1,2,\ldots,N\}$. 
We will recall this expression while we derive the $k$-separability condition 
which is suitable for all partitions. 
\section{Criteria for non-$k$-separability}
\label{sec3}
The separability condition which is given in terms of density matrix elements is found 
useful in detecting genuine multipartite entanglement and nonseparability of multipartite states \cite{guhne2010,gao2011,seev2008}. 
Inspired by this, we derive our $k$-separability condition by using the density matrix elements. 
The conditions given in this paper are applicable for a class of Dicke states with arbitrary excitations and $N$-qudit W states.  
To begin, we present the criterion to identify the non-$k$-separability of $N$-qubit states. \\

\noindent{\bf Criterion 1} : Let $\rho = (\rho_{i,j})_{2^N\times 2^N}$ be a $k$-separable $N$-qubit state.   
Then its density matrix elements fulfill

\small
\begin{align}
&\sum_{p_1=m-1}^{N-1} \sum_{p_2=m-2}^{N-2} \ldots \sum_{p_{m-1}=1}^{N-(m-1)}\sum_{p_m=0}^{N-m} 
\sum_{q_1=m-1}^{N-1} \sum_{q_2=m-2}^{N-2} \ldots \sum_{q_{m-1}=1}^{N-(m-1)}\sum_{q_m=0}^{N-m}\nonumber\\ 
& \qquad|\rho_{{2^{p_1}+2^{p_2}+\ldots+2^{p_m}+1},{2^{q_1}+2^{q_2}+\ldots+2^{q_m}+1}}| \nonumber\\
&\leq \sum_{i_1=0}^{N-(m-1)} \sum_{i_2=1}^{N-(m-2)} \ldots  \sum_{i_{m-1}=m-2}^{N-1}  \sum_{j_1=0}^{N-(m+1)} \sum_{j_2=1}^{N-m} \ldots \sum_{j_{m+1}=m}^{N-1}\nonumber\\ 
&\sqrt{\rho_{{2^{i_1}+2^{i_2}+\ldots+2^{i_{m-1}}+1},{2^{i_1}+2^{i_2}+\ldots+2^{i_{m-1}}+1}} 
\rho_{{2^{j_1}+2^{j_2}+\ldots+2^{j_{m+1}}+1},{2^{j_1}+2^{j_2}+\ldots+2^{j_{m+1}}+1}}} \nonumber \\
& +\left(\frac{N-k}{2}\right) \sum_{r_1=0}^{N-m} \sum_{r_2=1}^{N-(m-1)}\ldots \sum_{r_{m-1}=m-2}^{N-2} 
\sum_{r_{m}=m-1}^{N-1} \rho_{{2^{r_1}+2^{r_2}+\ldots+2^{r_m}+1},{2^{r_1}+2^{r_2}+\ldots+2^{r_m}+1}}. \label{dik} \qquad  
\end{align}
\normalsize
Here $p_k, q_k, r_k \in \{ 0,1,\ldots,N-1\}$, $1\leq k\leq m$, 
$i_r \in \{ 0,1,\ldots,N-1\}$, $1\leq r\leq m-1$, $j_s \in \{ 0,1,\ldots,N-1\}$, $1\leq s\leq m+1$. 
$p_m < p_{m-1} < \cdots < p_2 < p_1$, $q_m < q_{m-1} < \cdots < q_2 < q_1$, 
$q_m\geq p_m, q_{m-1}\geq p_{m-1},\ldots,q_1\geq p_1$, $(p_1,p_2,\ldots,p_{m-1},p_m)\neq (q_1,q_2,\ldots,q_{m-1},q_m)$.  
The set $\{p_1,p_2,\ldots,p_m\} \cap \{q_1,q_2,\ldots,q_m\}$ has exactly $(m-1)$ elements. 
$i_1 < i_2 < \cdots < i_{m-1}$, $j_1 < j_2 < \cdots < j_{m+1}$, 
$\{i_1,i_2,\ldots,i_{m-1}\} \cap \{j_1,j_2,\ldots,j_{m+1}\}$ $= \{i_1,i_2,\ldots,i_{m-1}\}$
and $r_1 < r_2 < \cdots < r_m$. 
 
An $N$-qubit state $\rho$ which violates the inequality (\ref{dik}), is a non-$k$-separable $N$-qubit state 
and if it violates the inequality (\ref{dik}) for $k=2$, then $\rho$ is a non-$2$-separable $N$-qubit state or  
a genuinely $N$-qubit entangled state. 
We have imposed certain parameter constraints in the condition in order to make the condition suitable for non-$k$-separable 
Dicke class of states with $m$ excitations.  For example, equation (\ref{dik}) 
for the $4$-qubit state with $2$ excitations read as 
\begin{align}  
\left. \begin{array}{r}
\rho_{4,6}+\rho_{4,7}+\rho_{4,10}\\
+\rho_{4,11}+\rho_{6,7}+\rho_{6,10}\\
+\rho_{6,13}+\rho_{7,11}+\rho_{7,13}\\
+\rho_{10,11}+\rho_{10,13}+\rho_{11,13}
\end{array} \right\} \leq 
\left\{ \begin{array}{l}
\sqrt{\rho_{2,2}\rho_{8,8}}+\sqrt{\rho_{3,3}\rho_{8,8}}+\sqrt{\rho_{2,2}\rho_{12,12}} \\
+\sqrt{\rho_{3,3}\rho_{12,12}}+\sqrt{\rho_{5,5}\rho_{8,8}}+\sqrt{\rho_{2,2}\rho_{14,14}} \\
+\sqrt{\rho_{5,5}\rho_{14,14}}+\sqrt{\rho_{3,3}\rho_{15,15}}+\sqrt{\rho_{5,5}\rho_{15,15}} \\ 
+\sqrt{\rho_{9,9}\rho_{12,12}}+\sqrt{\rho_{9,9}\rho_{14,14}}+\sqrt{\rho_{9,9}\rho_{15,15}} \\
+\left(\frac{4-k}{2}\right) \Big(\rho_{4,4} + \rho_{6,6} +\rho_{7,7} +\rho_{10,10}\\
\quad+\rho_{11,11}+\rho_{13,13}\Big)
\end{array} \right.. \qquad
\end{align}

In the following, we present another criterion which is applicable for a class of $N$-qudit W states \cite{gabr2010}. \\

\noindent{\bf Criterion 2} : Let $\rho = (\rho_{i,j})_{d^N\times d^N}$ be a $k$-separable $N$-partite state.   
Then its density matrix elements fulfill
\begin{align}
\sum_{P,Q} &|\rho_{\sum_{i=1}^N p_i~d^{d-i}+1,\sum_{j=1}^N q_j~d^{d-j}+1}| \nonumber\\
&\leq \sum_{R,S} \sqrt{\rho_{\sum_{i=1}^N r_i~d^{d-i}+1,\sum_{i=1}^N r_i~d^{d-i}+1} \rho_{\sum_{j=1}^N s_j~d^{d-j}+1,\sum_{j=1}^N s_j~d^{d-j}+1}} \nonumber \\
& \quad+\left(\frac{N-k}{2}\right) \sum_{T} \rho_{\sum_{i=1}^N t_i~d^{d-i}+1,\sum_{i=1}^N t_i~d^{d-i}+1}. \label{highw}
\end{align}
Here $P=\{p_1,p_2,\ldots,p_N\}$, $Q=\{q_1,q_2,\ldots,q_N\}$, $R=\{r_1,r_2,\ldots,r_N\}$, $S=\{s_1,s_2,\ldots,s_N\}$, $T=\{t_1,t_2,\ldots,t_N\}$ 
and $p_i,q_j,r_i,s_j,t_i \in \{0,1,\ldots,(d-1)\}$, $i,j\in\{1,2,\ldots,N\}$. 
The set $\{i\in\{1,2,\ldots,N\} : p_i=q_i\}$, $\{i\in\{1,2,\ldots,N\} : r_i=s_i\}$ has exactly $(N-2)$ elements. 
If $i\neq j$, then $p_i\neq p_j$ and $q_i\neq q_j$. 
If $p_i\neq q_i$, $p_j\neq q_j$ and $i<j$, then $p_i<q_i$ and $p_j>q_j$. 
If $r_i\neq s_i$ and $r_j\neq s_j$, then $r_i=r_j$, $s_i=s_j$ and $r_i<s_i$, $r_j<s_j$. If $i\neq j$, then $t_i \neq t_j$. 
We should always consider $d=N$, in order to be the state $|W_N^d\rangle$ (\ref{nqdw}). 

An $N$-partite state $\rho$ which violates the inequality (\ref{highw}), 
is a non-$k$-separable $N$-partite state 
and if it violates the inequality (\ref{highw}) for $k=2$, 
then $\rho$ is a genuinely multipartite entangled state. 
Equation (\ref{highw}) for the $3$-qutrit state read as 
\begin{align}  
\left. \begin{array}{r}
\rho_{6,8}+\rho_{6,12}+\rho_{6,22}\\
+\rho_{8,16}+\rho_{8,20}+\rho_{12,16}\\
+\rho_{12,20}+\rho_{16,22}+\rho_{20,22}
\end{array} \right\} \leq 
\left\{ \begin{array}{l}
\sqrt{\rho_{5,5}\rho_{9,9}}+\sqrt{\rho_{3,3}\rho_{15,15}}+\sqrt{\rho_{4,4}\rho_{24,24}} \\
+\sqrt{\rho_{7,7}\rho_{17,17}}+\sqrt{\rho_{2,2}\rho_{26,26}} \\
+\sqrt{\rho_{10,10}\rho_{18,18}}+\sqrt{\rho_{11,11}\rho_{21,21}}\\ 
+\sqrt{\rho_{13,13}\rho_{25,25}}+\sqrt{\rho_{19,19}\rho_{23,23}} \\ 
+\left(\frac{3-k}{2}\right) \Big(\rho_{6,6} + \rho_{8,8} +\rho_{12,12} \\ 
\quad +\rho_{16,16}+\rho_{20,20}+\rho_{22,22}\Big)
\end{array} \right..
\end{align} 

To construct the conditions (\ref{dik}) and (\ref{highw}), we choose certain off-diagonal elements in the density matrix of Dicke states with $m$ excitations 
and higher dimensional W states and collect their corresponding diagonal elements by considering all possible partitions of $k$-separable $N$-partite states.  
The term $\frac{N-k}{2}$ which appear in the inequalities (\ref{dik}) and (\ref{highw}) decides the non-$k$-separability of $N$-partite states. 
The inequalities given above can be verified in the same manner as the Theorem 3 in \cite{gao2011} was proved. 
However, deriving the non-$k$-separability conditions for the above two classes of states 
is not a simple linear combination of the conditions established for the qubit case.  
In the present case we need to consider several possibilities and impose several constraints due to different excitations of the Dicke class of states and $N$-qudit W states. 
In Appendix, we present the proof of the inequality (\ref{dik}).   
The proof of the second criterion is analog to the proof given for the criterion 1.    
\section{Illustration with examples} 
\label{sec4}
In this section, we illustrate the performance of our criteria 1 and 2 in different situations. 
We first analyze the non-$k$-separability of $N$-qubit Dicke state added with white noise. 
We then investigate the genuine multipartite entanglement and nonseparability of Dicke states added with white noise under different excitations.   
We also study the noise robustness of our criterion 1 for $N$-qubit Dicke states with $m$ excitations added with white noise by varying $m$ and $k$. 
We then analyze the non-$k$-separability of $N$-partite W state added with white noise and study 
the noise robustness of our criterion 2 for this state by varying $k$. \\

\noindent 1. Let us consider a $N$-qubit Dicke state with $m$ excitations added with white noise \cite{hube2011}, 
\begin{align}
\label{rhod} \rho_D = (1-p) |D_m^N \rangle\langle D_m^N | + p \frac{\mathbb{I}}{2^N}, 
\end{align}
where $\mathbb{I}$ is the identity operator. 
Applying the condition (\ref{dik}) on (\ref{rhod}), we can obtain the following general function, namely
\begin{align}
\label{gamma} \gamma_k^{N,m} = \frac{\left((N-m)\frac{m}{2}\left(\frac{p}{2^N}\right) + \left(\frac{N-k}{2}\right)\left(\frac{1-p}{\left({N\atop m}\right)}+\frac{p}{2^N}\right)\right)\left({N\atop m}\right)}{(N-m)\frac{m}{2}(1-p)}. 
\end{align}
\begin{figure}[h!]
\centering
  \includegraphics[width=0.48\textwidth]{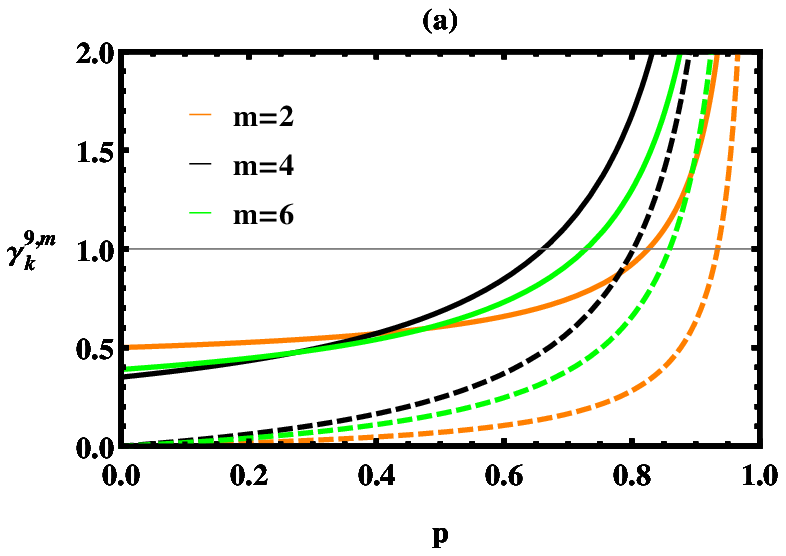} \quad \includegraphics[width=0.48\textwidth]{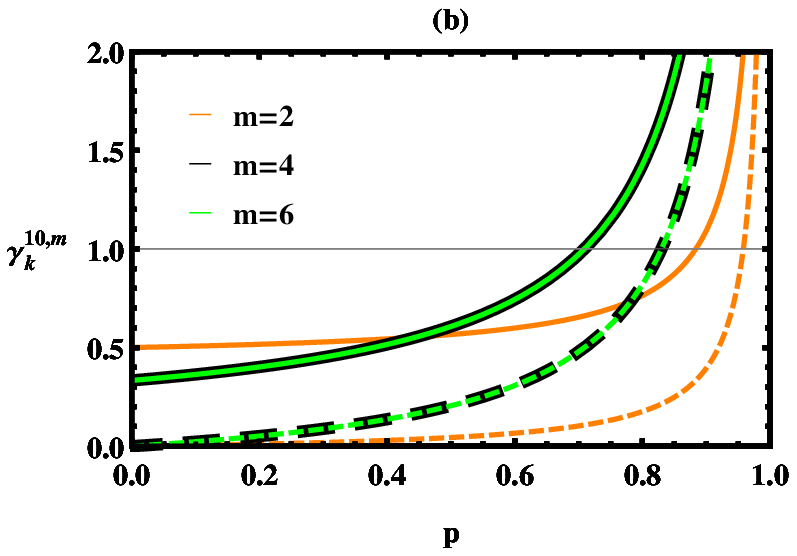} \\
      \includegraphics[width=0.48\textwidth]{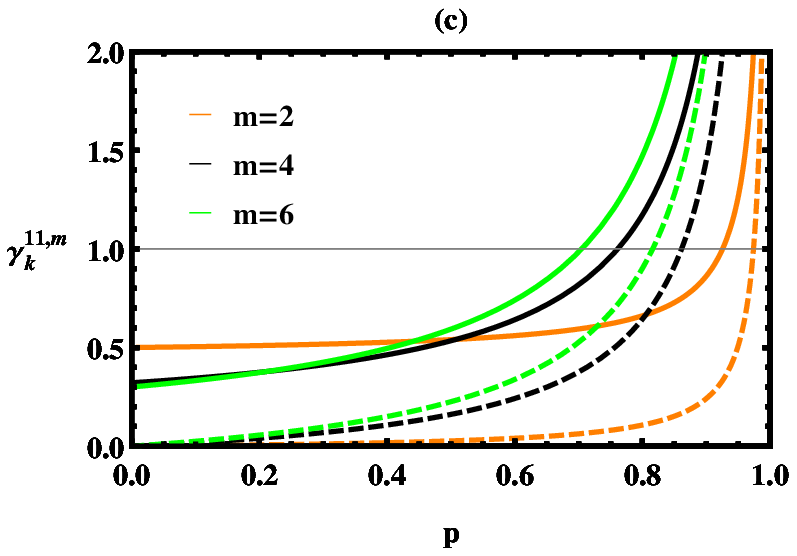}  
  \caption{The ranges of genuinely multipartite entanglement (solid curve) and nonseparability (dashed curve) of $N$-qubit Dicke state added with white noise with excitations $m=2$, $m=4$ and $m=6$. The ranges are covered in the order for (a) $N=9$, $m=2 > m=6 > m=4$, (b) $N=10$, $m=2 > (m=4) = (m=6)$ and (c) $N=11$, $m=2>m=4>m=6$.} \label{9qu}
\end{figure}
When the state $\rho_D$ obeys inequality $\gamma_k^{N,m} < 1$, for a given value of $k$ and for the parameter $(p)$ range, 
then the state is non-$k$-separable. 
For example, a $4$-qubit Dicke state with 2 excitations added with white noise, 
the criterion 1 detects this state as genuinely multipartite entangled for $p<0.471$.  
This result matches with the one presented in \cite{hube2011}. 
For the phased Dicke state \cite{chiu2010} added with white noise we observe that 
our criterion acts as strong as the criterion given in Ref.\cite{hube2011}. 
The function $\gamma_k^{N,m}$ can also be employed to identify the 
non-$k$-separability $(2 \leq k\leq N)$ for different $N$ qubit states with $m$ excitations.  

Next we investigate the genuine multipartite entanglement (GME) and nonseparability (NS) 
of the state $\rho_D$ with different excitations in $|D_m^N \rangle$. 
For this purpose, we analyze the function $\gamma_k^{N,m}$ for $N=9,10$ and $11$ with $m=2,4,6$ and $k=2$ and $N$ and depict 
the outcome in figures \ref{9qu}(a), \ref{9qu}(b) and \ref{9qu}(c) respectively. 
In Fig.\ref{9qu}(a), for $N=9$, the genuine multipartite entanglement range 
(range of solid curves with $\gamma_2^{9,m}<1$)  
of a $9$-qubit state $\rho_D$ with different excitations are observed in the following order,  
that is $m=2 > m=6 > m=4$. 
Similarly, the nonseparability 
(range of dashed curves with $\gamma_9^{9,m}<1$)  
of a $9$-qubit state $\rho_D$ with different excitations are also found in the same order, that is $m=2 > m=6 > m=4$. 
From the graphs we infer that the case $m=2$ ($|D_2^9 \rangle$ in $\rho_D$) 
is more entangled than the cases $m=6$ ($|D_6^9 \rangle$ in $\rho_D$) and $m=4$ ($|D_4^9 \rangle$ in $\rho_D$). 
Among the later, the case $m=6$ is more entangled than $m=4$. 
In Fig.\ref{9qu}(b), we display the outcome for $N=10$. 
Here we observe that the GME and NS range appear as  
$m=2 > (m=4) = (m=6)$. 
For $N=11$, as we see in Fig.\ref{9qu}(c), the range covered by the order turns out that $m=2>m=4>m=6$.   
The same order $m=2>m=4>m=6$ is being preserved even for higher values of $N$ $(N>11)$. 
Let $m_i$ and $m_j$ be the $i$ and $j$ excitations with $i,j=2,3,\ldots,m$. 
If $m_i+m_j>N$, the range of genuine multipartite entanglement and nonseparability  
of $N$-qubit state $\rho_D$ with different excitations 
differ from the order $m_2>m_3>\ldots>m_m$ or $m_2<m_3<\ldots<m_m$.  
When $m_i+m_j=N$ the range correspond to the excitations $m_i$ and $m_j$ 
overlap with each other.  
If we consider any two maximum excitations $m_i$ and $m_j$ with $m_i+m_j<N$, then 
the GME and NS range for different excitations 
will emerge as $m_2>m_3\ldots>m_m$. 
In general, the entanglement of the state (\ref{rhod}) decreases when we increase the excitation.    
\begin{figure}[h!]
\centering
  \includegraphics[width=0.47\textwidth]{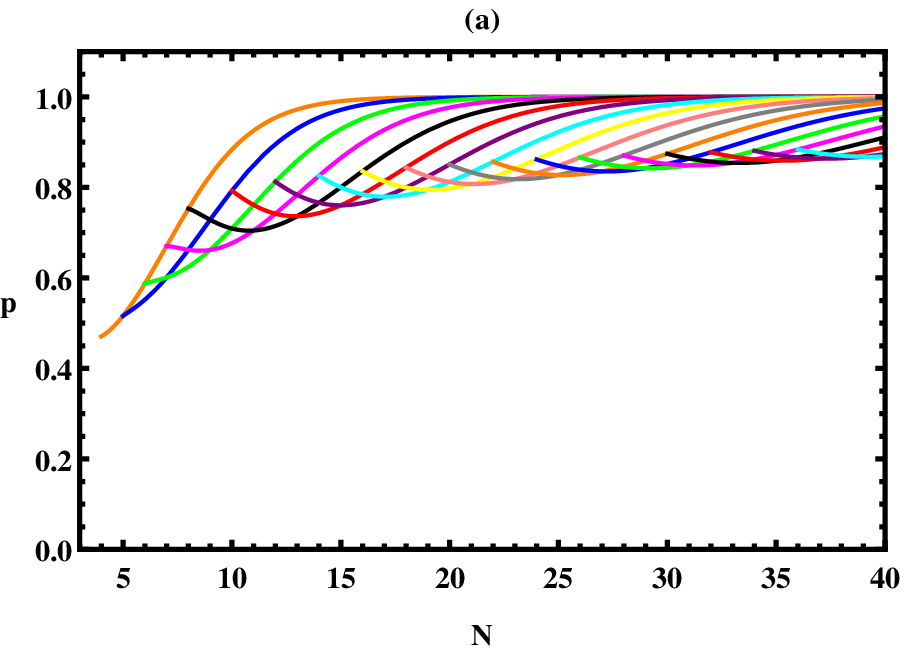} \quad \includegraphics[width=0.47\textwidth]{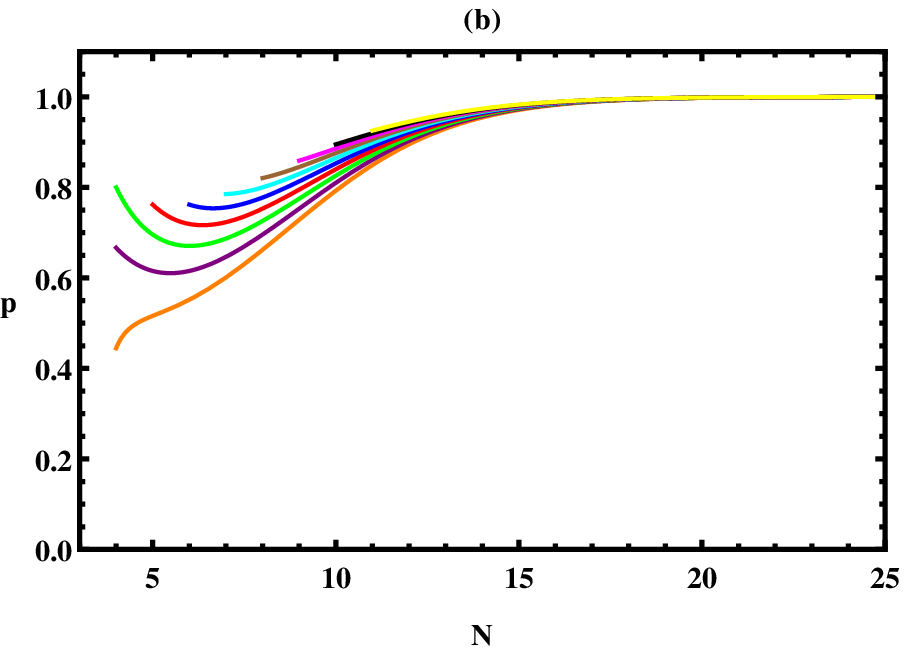}
  \caption{The white noise tolerance for (a) $N$ number of qubits with different excitations, where $m$ varies from $2$ to $20$ in ascending order (from left to right) and 
(b) $N$ number of qubits with different $k$, where $k$ varies from $2$ to $11$ in ascending order (from bottom to top).} 
\label{wnt1}
\end{figure}

To study the white noise tolerance of criterion 1 for $N$-qubit Dicke states with $m$ excitations added with white noise, we derive  
the following expression from Eq.(\ref{gamma}), that is  
\begin{align}
p<\frac{(N-m)\frac{m}{2} - \left(\frac{N-k}{2}\right)}{\left(\frac{N-m}{2^N}\right)\frac{m}{2}\left({N\atop m}\right) - \left(\frac{N-k}{2}\right)+\left(\frac{N-k}{2^{N+1}}\right) \left({N\atop m}\right) + (N-m)\frac{m}{2}},  \label{wnr}
\end{align}
and plot this function for various number of N-qubits with different excitations. 
The outcome is depicted in Fig.\ref{wnt1}(a).  
The figure reveals that for large number of systems the white noise tolerance increases rapidly and reaches the value $1$. 
For example, when $k=2$ and $N\geq 20$, there will be no effect of white noise on Dicke state with $2$ excitations according to our criterion and it remains an entangled state. 
This result also matches with the one reported in \cite{hube2011}. 
Next we fix $m=3$ in (\ref{wnr}) and vary the number of qubits and $k$. 
In this case also the white noise tolerance approaches the value $1$ for 
large $N$ $(N>20)$ regardless the value of $k$, 
which can be seen in Fig.\ref{wnt1}(b). \\

\begin{figure}[t]
\centering
 \includegraphics[width=0.43\textwidth]{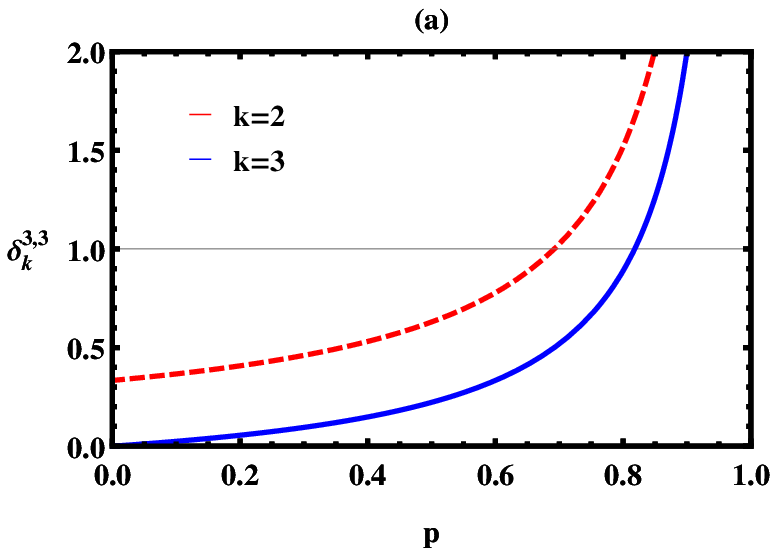} \quad \includegraphics[width=0.43\textwidth]{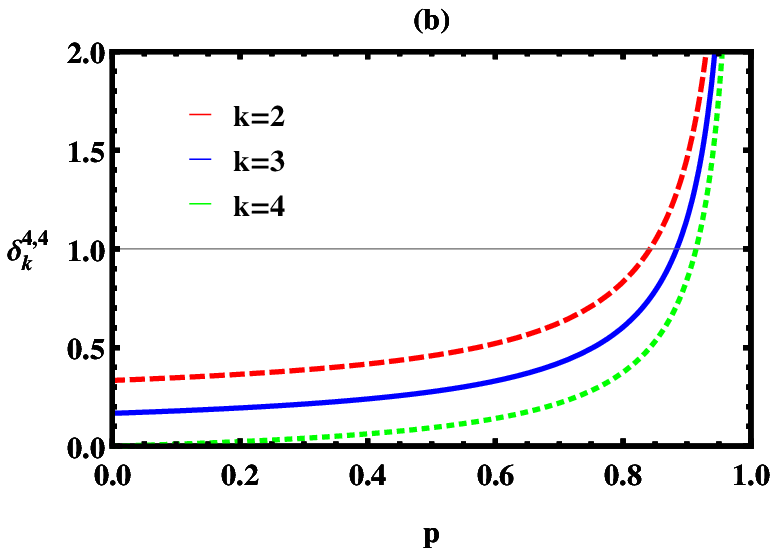} 
 \caption{Non-$k$-separability of (a) $3$-qutrit W state added with white noise and (b) $4$-dimensional $4$-partite W state added with white noise.} \label{f6}
\end{figure}

\noindent 2.  Let us consider a $N$-qudit W state added with white noise \cite{gabr2010}, 
\begin{align}
\label{rhow} \rho_W = (1-p) |W_N^d \rangle\langle W_N^d | + p \frac{\mathbb{I}}{d^N}, 
\end{align}
where $\mathbb{I}$ is the identity operator. 
Applying the condition (\ref{highw}) on (\ref{rhow}), we can obtain the following general function, namely 
\begin{align}
\label{delta} \delta_k^{N,d} = \frac{N!~p}{d^N~(1-p)}+\frac{2~(N-k)}{N~(N-1)} + \frac{2~(N-k)~(N-2)!~p}{d^N~(1-p)}.  
\end{align} 
When the state $\rho_W$ obeys inequality $\delta_k^{N,d} < 1$, for a given value of $k$ and for the parameter $(p)$ range, 
then the state is non-$k$-separable. 
For example, a $3$-qutrit W state added with white noise, 
the criterion 2 detects the state (\ref{rhow}) as genuinely multipartite entangled for $p < 0.693$. 
For a $3$-qutrit W state added with white noise, the criterion come from the linear combinations 
of all off-diagonal elements detects the state as genuinely multipartite entangled for $p < 0.445$ only. 
We mention here that the obtained result also differs from the one come from the linear combination 
of off-diagonal elements, whereas the present criterion has larger detection range. 
Hence this condition acts as a strong condition for this class of states. 
To illustrate the non-$k$-separability, let us consider two cases, namely (i) $3$-qutrit ($n=3$ and $d=3$) and 
(ii) $4$-dimensional $4$-partite ($n=4$ and $d=4$) states in (\ref{rhow}). For these two cases, equation (\ref{delta}) gives us  
$\delta_k^{3,3}=\frac{2~p}{9~(1-p)}+\frac{(3-k)}{3} + \frac{2~p~(3-k)}{27~(1-p)}$ and 
$\delta_k^{4,4}=\frac{3~p}{32~(1-p)}+\frac{(4-k)}{6}+\frac{(4-k)~p}{24~(1-p)}$, respectively.  
We plot these two functions for various $k$ $(2\leq k\leq N)$ values and depict the outcome in Figs.\ref{f6}(a) and \ref{f6}(b) respectively.  
In these two figures, the range covered by $\delta_k^{N,d} < 1$ explores the non-$k$-separability. 
\begin{figure}[ht]
\centering
 \includegraphics[width=0.38\textwidth]{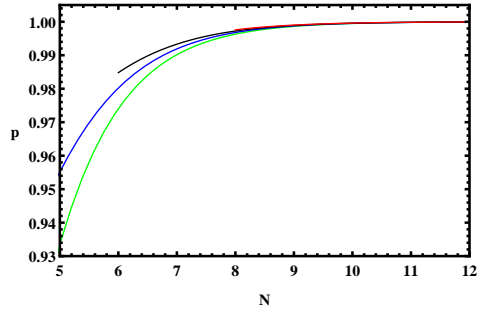}
 \caption{The white noise tolerance for $N$ number of qudits with different $k$, where $k = 2,4,6,8$ (varies from bottom to top).} \label{f8} 
\end{figure}
To study the white noise tolerance for $N$-qudit W state added with white noise, we derive  
the following expression from Eq.(\ref{delta}), that is  
\begin{align}
p< \frac{d^N~(2k + (N-3)N)}{d^N~(2k+(N-3)N) + (N^2 + N - 2k)~N!},  \label{wnwd}
\end{align}
and plot this function for various number of $N$-qudits (where $d=N$). The outcome is given in Fig.\ref{f8}.  
In this case also the white noise tolerance approaches the value $1$ for $N\geq 12$ regardless the value of $k$, 
which can be seen from Fig.\ref{f8}. 

\section{Experimental feasibility}
\label{sec6} 
The conditions (\ref{dik}) and (\ref{highw}) which are formulated in terms of density matrix elements 
may cumbersome to read. However, they can be easily determined from the expectation values of local observables  \cite{seev2008,guhne2003,guhn2007,gao2010}, as we see below. 
To determine the off-diagonal and diagonal elements that appear in the expression (\ref{dik}) 
we present local observables in terms of Pauli operators \cite{hube2011}. 
The off-diagonal elements can be determined by measuring the observables $O$ and $\tilde{O}$ given below, 
that is  
\begin{subequations}
\label{off}
\begin{align}
O = &\frac{\sigma_x^i \sigma_x^j}{2^{N-1}} \otimes \left((-1)^{ \displaystyle\sum_i \gamma_{a_i,b_i}} \sum_{s=0}^{N-2} \sum_p P_p(\mathbb{I}^{\otimes s} \sigma_z^{\otimes N-2-s})^{k_1k_2\ldots k_m} \right) \nonumber \\
 & + \frac{\sigma_y^i \sigma_y^j}{2^{N-1}} \otimes \left((-1)^{ \displaystyle\sum_i \gamma_{a_i,b_i}} \sum_{s=0}^{N-2} \sum_p P_p(\mathbb{I}^{\otimes s} \sigma_z^{\otimes N-2-s})^{k_1k_2\ldots k_m} \right),  \\
\tilde{O} =& \frac{\sigma_x^i \sigma_y^j}{2^{N-1}} \otimes \left((-1)^{ \displaystyle\sum_i \gamma_{a_i,b_i}} \sum_{s=0}^{N-2} \sum_p P_p(\mathbb{I}^{\otimes s} \sigma_z^{\otimes N-2-s})^{k_1k_2\ldots k_m} \right) \nonumber \\
&- \frac{\sigma_y^i \sigma_x^j}{2^{N-1}} \otimes \left((-1)^{ \displaystyle\sum_i \gamma_{a_i,b_i}} \sum_{s=0}^{N-2} \sum_p P_p(\mathbb{I}^{\otimes s} \sigma_z^{\otimes N-2-s})^{k_1k_2\ldots k_m} \right),  
\end{align}
\end{subequations}
where $k_1,k_2,\ldots,k_m = \{ 1,2,,\ldots,N\}$, $k_1\neq k_2 \neq\ldots\neq k_m$, $i,j = \{ 1,2,,\ldots,N\}$, $i<j$,  
$\{i,j\} \neq \{ k_1,k_2,\ldots,k_m \}$ and $\sum_p P_p(\cdot)$ denotes sum over all possible permutations. 
The operators $O$ and $\tilde{O}$ determine the real and imaginary parts of the off-diagonal elements respectively. 
Similarly, the diagonal elements can be determined by measuring the observable $D$ through the expression  
\begin{align}
D = \frac{1}{2^N}\left((-1)^{ \displaystyle\sum_i \gamma_{a_i,b_i}} \sum_{s=0}^{N} 
\sum_p P_p(\mathbb{I}^{\otimes s} \sigma_z^{\otimes N-s}) \right). 
\label{dia}
\end{align}

The term $\sum_i \gamma_{a_i,b_i}$ determines the sign $(+$ or $-)$ of each term that appear 
in the right hand side of expressions (\ref{off}) and (\ref{dia}). 
To evaluate the term $\gamma_{a_i,b_i}$ 
(i) we pick up an element (which is to be measured) from the density matrix and write its basis,  
(ii) we expand the respective operator (see equations (\ref{off}) and (\ref{dia})) for the chosen element from 
the $N$-qubit density matrix,   
(iii) we consider all the $|1\rangle_i\langle 1|$ states of the $i^\textrm{th}$ subsystem(s) in the basis which  
we call as $a_i$ $(a_i=|1\rangle_i\langle 1|)$,  
(iv) for each local observables in the operator, we consider a Pauli operator ($\sigma_z^i$ or $\mathbb{I}^i$) 
that acts on the $i^\textrm{th}$ subsystem which we call as $b_i$ ($b_i = \sigma_z^i$ or $\mathbb{I}^i$),   
(v) we then match this $|1\rangle_i\langle 1|$ with $i^\textrm{th}$ operator for some $i$ and  
assign the value $1$ for the case $\gamma_{a_i,b_i} = \gamma_{|1\rangle_i\langle 1|,\sigma_z^i}$ and 
$0$ for the case $\gamma_{a_i,b_i} = \gamma_{|1\rangle_i\langle 1|,\mathbb{I}^i}$ respectively    
(hence $\gamma_{a_i,b_i}$ can be either $0$ or $1$) and 
(vi) we consider all possible $|1\rangle_i\langle 1|$ with their corresponding operators 
$\sigma_z^i$ or $\mathbb{I}^i$ that appear in each local observables  
and add all the obtained values of $\gamma_{a_i,b_i}$ to get the net value of $\sum_i\gamma_{a_i,b_i}$ 
for each term in the expressions (\ref{off}) and (\ref{dia}). 

To illustrate the above said procedure, let us consider a diagonal element, say for example $\rho_{7,7}$, of 
a $5$-qubit state whose basis is given by $|00110\rangle\langle 00110|$. 
The observable $D$ corresponds to this diagonal element reads (see equation (\ref{dia})),
\small
\begin{align}
D =&\frac{(-1)^{\sum_i \gamma_{a_i,b_i}}}{32} \Big(\sigma_z\sigma_z\sigma_z\sigma_z\sigma_z + \mathbb{I}\sigma_z\sigma_z\sigma_z\sigma_z + \sigma_z\mathbb{I}\sigma_z\sigma_z\sigma_z + \sigma_z\sigma_z\mathbb{I}\sigma_z\sigma_z + \sigma_z\sigma_z\sigma_z\mathbb{I}\sigma_z \nonumber\\ 
&+ \sigma_z\sigma_z\sigma_z\sigma_z\mathbb{I} + \mathbb{I}\mathbb{I}\sigma_z\sigma_z\sigma_z + \mathbb{I}\sigma_z\mathbb{I}\sigma_z\sigma_z + \mathbb{I}\sigma_z\sigma_z\mathbb{I}\sigma_z + \mathbb{I}\sigma_z\sigma_z\sigma_z\mathbb{I} + \sigma_z \mathbb{I}\mathbb{I}\sigma_z\sigma_z \nonumber\\ 
&+ \sigma_z \mathbb{I}\sigma_z\mathbb{I}\sigma_z + \sigma_z \mathbb{I}\sigma_z\sigma_z \mathbb{I} + \sigma_z \sigma_z\mathbb{I}\mathbb{I}\sigma_z + \sigma_z \sigma_z\mathbb{I}\sigma_z\mathbb{I} + \sigma_z \sigma_z\sigma_z\mathbb{I}\mathbb{I} + \mathbb{I}\mathbb{I}\mathbb{I}\sigma_z\sigma_z \nonumber\\
&+ \mathbb{I}\mathbb{I}\sigma_z\mathbb{I}\sigma_z + \mathbb{I}\mathbb{I}\sigma_z\sigma_z\mathbb{I} + 
\mathbb{I}\sigma_z\mathbb{I}\mathbb{I}\sigma_z + \mathbb{I}\sigma_z\mathbb{I}\sigma_z\mathbb{I} + 
\mathbb{I}\sigma_z\sigma_z\mathbb{I}\mathbb{I} + \sigma_z\mathbb{I}\mathbb{I}\mathbb{I}\sigma_z + \sigma_z\mathbb{I}\mathbb{I}\sigma_z\mathbb{I} \nonumber\\
&+ \sigma_z\mathbb{I}\sigma_z\mathbb{I}\mathbb{I} + \sigma_z\sigma_z\mathbb{I}\mathbb{I}\mathbb{I} + \mathbb{I}\mathbb{I}\mathbb{I}\mathbb{I}\sigma_z + \mathbb{I}\mathbb{I}\mathbb{I}\sigma_z\mathbb{I} + \mathbb{I}\mathbb{I}\sigma_z\mathbb{I}\mathbb{I} 
+ \mathbb{I}\sigma_z\mathbb{I}\mathbb{I}\mathbb{I} + \sigma_z\mathbb{I}\mathbb{I}\mathbb{I}\mathbb{I} 
+ \mathbb{I}\mathbb{I}\mathbb{I}\mathbb{I}\mathbb{I} \Big). \qquad \label{d5q}
\end{align}
\normalsize
Let us consider a specific term now, say for example the  $8^{\textrm{th}}$ term in the above equation (\ref{d5q}), that is 
$(-1)^{\sum_i \gamma_{a_i,b_i}}\mathbb{I}^1\sigma_z^2\mathbb{I}^3\sigma_z^4\sigma_z^5/32$ 
and compare it with 
$|0\rangle_1\langle 0|\otimes|0\rangle_2\langle 0|\otimes|1\rangle_3\langle1|\otimes|1\rangle_4\langle1|\otimes|0\rangle_5\langle 0|$. 
Doing so, we find $|1\rangle_3\langle1| =  \mathbb{I}^3$. So we assign $\gamma_{a_i,b_i} =0$.    
For the fourth subsystem, we find $|1\rangle_4\langle1| = \sigma_z^4$ and so we assign $\gamma_{a_i,b_i} =1$. 
Therefore the term $(-1)^{\sum_i \gamma_{a_i,b_i}}$$\mathbb{I}\sigma_z\mathbb{I}\sigma_z\sigma_z/32$ yields 
$(-1)^{(0+1)}$$\mathbb{I}\sigma_z\mathbb{I}\sigma_z\sigma_z/32$ (or) $-\mathbb{I}\sigma_z\mathbb{I}\sigma_z\sigma_z/32$ 
for the basis $|00110\rangle\langle 00110|$. 
In this manner, we evaluate the sum $\sum_i \gamma_{a_i,b_i}$ for each and every term 
that appear on the right hand side of the expression (\ref{d5q}) and 
fix the sign $(+~\textrm{or}~-)$ of each term and construct the operator (\ref{dia}). 
A similar procedure is also followed for the off-diagonal elements with the help of (\ref{off}). 

The off-diagonal elements that appear in the left hand side of (\ref{dik}) can be 
determined from $(N-m)\frac{m}{2}\left(N\atop m\right)2^{N-1}$ number of local observables. 
To determine the diagonal elements appear in the right hand side of the inequality (\ref{dik}), 
for the $N$-qubit state, one requires $2^N$ local observables. 
Therefore, one requires totally $(N-m)\frac{m}{2}\left(N\atop m\right)2^{N-1}+2^N$ number of local observables to evaluate the criterion 1.  
It is sufficient to identify the non-$k$-separability of $N$-qubit state with $m$ excitations. 
For higher excitations, we noticed that the number of off-diagonal elements of density matrix will be less and so 
the required local observables which are to be measured should also be less. 

In a similar manner, to determine the off-diagonal and diagonal elements 
that appear in the expression (\ref{highw}) 
we present local observables in terms of generalized Gell-Mann matrices \cite{bert2007,zhao2014}. 
We consider the generalized Gell-Mann matrices (GGM) for $d$-dimensional vector space
\begin{subequations}
\begin{align}
\lambda^{jk} =& |j\rangle\langle k| + |k\rangle\langle j|, \qquad\qquad\qquad\qquad\qquad\qquad\quad 0\leq j<k \leq d-1, \qquad \\
\mu^{jk} =&-i |j\rangle\langle k| + i |k\rangle\langle j|,  \qquad\qquad\qquad\qquad\qquad\quad 0\leq j<k \leq d-1, \qquad 
\end{align}
\end{subequations}

\renewcommand{\theequation}{\arabic{equation}c}
\begin{align}
\setcounter{equation}{17}
\eta^l =& \sqrt{\frac{2}{(l+1)(l+2)}} \left(\sum_{j=0}^l |j\rangle\langle j| - (l+1) |l+1\rangle\langle l+1 | \right), 0\leq l\leq d-2. \quad
\end{align}
Here totally, we have $d^2-1$ GGM which are Hermitian and traceless. 
The operators $|j\rangle\langle k|$ and $|j\rangle\langle j|$ with $j,k=0,1,\ldots,(d-1)$,  
can also be expressed in terms of GGM \cite{zhao2014}, that is 
\renewcommand{\theequation}{\arabic{equation}}
\begin{subequations}
\begin{align}
|j\rangle\langle k| =& \frac{1}{2} \left(\lambda^{jk} + i \mu^{jk} \right), \qquad\qquad\qquad\qquad\qquad\qquad \textrm{for}\quad j<k, \\
|j\rangle\langle k| =& \frac{1}{2} \left(\lambda^{jk} - i \mu^{jk} \right), \qquad\qquad\qquad\qquad\qquad\qquad \textrm{for}\quad j>k, \\
|j_i\rangle\langle j_i| =& -\sqrt{\frac{j_i}{2~(j_i+1)}} \eta^{j_i-1} + \sum_{m=0}^{d-j_i-2} \frac{1}{\sqrt{2~(j_i+m+1)~(j_i+m+2)}} \eta^{j_i+m} + \frac{1}{d} \mathbb{I}, \qquad  
\end{align} 
\end{subequations}
where $j_i = 0,1,\ldots,(d-1)$. The off-diagonal elements that appear in the expression (\ref{highw}) 
can be determined by measuring the observables $Q$ and $\tilde{Q}$ through the relations  
\begin{subequations}
\begin{align} 
Q =& \frac{1}{4} \left( \lambda_a^{jk} \lambda_b^{jk} + i \lambda_a^{jk}  \mu_b^{jk} - i \mu_a^{jk} \lambda_b^{jk} + \mu_a^{jk} \mu_b^{jk} \right) \otimes 
\left(|j_i\rangle\langle j_i|^{\otimes N-2}\right)_{k_1,\ldots,k_n}, \\ 
\tilde{Q} =& \frac{1}{4} \left( \lambda_a^{jk} \lambda_b^{jk} - i \lambda_a^{jk}  \mu_b^{jk} + i \mu_a^{jk} \lambda_b^{jk} + \mu_a^{jk} \mu_b^{jk} \right) \otimes 
\left(|j_i\rangle\langle j_i|^{\otimes N-2}\right)_{k_1,\ldots,k_n},
\end{align}
\end{subequations}
where $a,b = \{1,2,\ldots,N\}$, $a\neq b$, $k_1,k_2,\ldots,k_m = \{1,2,\ldots,N\}$, $k_i \neq k_j$ if $i\neq j$. 
The operators $Q$ and $\tilde{Q}$ determine the real and imaginary parts of the off-diagonal elements respectively. 
Similary, the diagonal elements that appear in the expression (\ref{highw}) can be determined by measuring the observable $D_d$ as   
\begin{align}
D_d = \bigotimes_{i=1}^N |j_i\rangle\langle j_i |. 
\end{align}
Therefore the off-diagonal elements appear in the left hand side of the inequality (\ref{highw}) of 
a $N$-qudit state 
can be determined from $\frac{2N~N!}{(N-2)!}$ local observables. 
To determine the diagonal elements appear in the right hand side of inequality (\ref{highw}) for the $N$-qudit state, one requires $d^N$ local observables. 
To evaluate the criterion 2 one requires totally $\frac{2N~N!}{(N-2)!}+d^N$ 
number of local observables. 
Thus our approach requires only fewer measurements when compare to the full quantum state tomography.  
\section{Conclusion}
\label{sec7}
We have proposed the necessary conditions to identify the non-$k$-separability in the Dicke class of states 
and $N$-qudit W class of states.   
To the authors knowledge goes this is the first non-$k$-separability criterion for 
the Dicke class of state with arbitrary excitations 
and strong non-$k$-separability criterion for a class of $N$-qudit W states.   
Using these criteria, we have demonstrated the genuine multiqubit entanglement and non-$k$-separability of 
$N$-qubit Dicke state added with white noise under different excitations and 
$N$-qudit W state added with white noise. 
For the above two mixed states, the white noise tolerance rapidly increases for large number of 
systems regardless the values of $m$ and $k$ and 
the white noise tolerance approaches 1 for $N\geq 12$ regardless the value of $k$. 
We then presented the local observables to determine the off-diagonal and 
diagonal density matrix elements in terms of Pauli operators and generalized Gell-Mann matrices.  
This in turn provides the experimental feasibility of our criteria.  
To identify the non-$k$-separability in $N$-qubit state with $m$ excitations and $N$-qudit state through our criteria one requires   
$(N-m)\frac{m}{2}\left(N\atop m\right)2^{N-1}+2^N$ and $\frac{2N~N!}{(N-2)!}+d^N$ number of local observables respectively. 
Therefore, our criteria are easily computable without optimization or eigenvalue evalution.  
\section*{Appendix : Proof of the Criterion (\ref{dik})} 
\label{proof}
\renewcommand{\theequation}{A.\arabic{equation}}
Let us consider an arbitrary pure $k$-separable $N$-qubit state 
\small
\begin{align}
\setcounter{equation}{0}
|\psi_{k\textrm{-sep}}\rangle =& |\psi_1\rangle_{x_1,x_2,\ldots,x_{N_1}} \otimes |\psi_2\rangle_{x_{N_1+1},x_{N_1+2},\ldots,x_{N_2}} \otimes \cdots \notag\\
& \otimes|\psi_{k-1}\rangle_{x_{N_{N-2}+1},x_{N_{N-2}+2},\ldots,x_{N_{N-1}}} \otimes |\psi_{k}\rangle_{x_{N_{N-1}+1},x_{N_{N-1}+2},\ldots,x_{N_N}}, \label{app1}
\end{align}
\normalsize
where $\{ x_1, x_2, \ldots, x_{N_N}\} = \{ 1,2, \ldots, N\}$ and 
$A_1=\{x_1,x_2,\ldots,x_{N_1}\}$, $A_2=\{x_{N_1+1},$ $x_{N_1+2},\ldots,x_{N_2}\}$,$\ldots$, 
$A_k=\{x_{N_{N-1}+1},x_{N_{N-1}+2},\ldots,x_{N_N}\}$. 
 
Let us consider $\{p_1,p_2,\ldots,p_m\}$  and $\{q_1,q_2,\ldots,q_m\}$ with the cases 
$p_1\in A_1$; $p_2\in A_2$;$\ldots$;$p_m\in A_{k-1}$; $q_1\in A_1$; $q_2\in A_2$;$\ldots$;$q_m\in A_{k-1}$,$\ldots$,  
$p_1\in A_1$; $p_2\in A_2$;$\ldots$; $p_m\in A_{k}$; $q_1\in A_1$; $q_2\in A_2$;$\ldots$;$q_m\in A_{k}$,$\ldots$,    
$p_1,p_2\in A_1$; $p_3,p_4\in A_2$;$\ldots$;$p_{m-1},p_m\in A_{k}$; $q_1,q_2\in A_1$; $q_3,q_4\in A_2$;$\ldots$;$q_{m-1},q_m\in A_{k}$,$\ldots$,  
$p_1,p_2,\ldots,$ $p_m\in A_k$; $q_1,q_2,\ldots, q_m\in A_k$  along with other constraints given in section \ref{sec3}. 
Then the density matrix element can be obtained as 
\small
\begin{align}
&|\rho_{{2^{p_1}+2^{p_2}+\cdots+2^{p_m}+1},{2^{q_1}+2^{q_2}+\cdots+2^{q_m}+1}}| \notag\\ 
&=\sqrt{\rho_{{2^{p_1}+2^{p_2}+\cdots+2^{p_m}+1},{2^{p_1}+2^{p_2}+\cdots+2^{p_m}+1}} \rho_{{2^{q_1}+2^{q_2}+\cdots+2^{q_m}+1},{2^{q_1}+2^{q_2}+\cdots+2^{q_m}+1}}} \notag\\
&\leq \frac{\rho_{{2^{p_1}+2^{p_2}+\cdots+2^{p_m}+1},{2^{p_1}+2^{p_2}+\cdots+2^{p_m}+1}}+\rho_{{2^{q_1}+2^{q_2}+\cdots+2^{q_m}+1},{2^{q_1}+2^{q_2}+\cdots+2^{q_m}+1}}}{2}.   \label{app2}
\end{align}
\normalsize 
The density matrix element for other cases like 
$\{p_1,p_2,\ldots,p_m\}\in A_1$; $\{q_1,q_2,\ldots,$ $q_{m-1}\}\in A_1$; $q_m\in A_2\cup\cdots\cup A_k$,$\ldots$, 
$\{p_1,p_2,\ldots,p_m\}\in A_k$; $\{q_1,q_2,\ldots,q_{m-1}\}\in A_k$; $q_m\in A_1\cup A_2\cup\cdots\cup A_{k-1}$,$\ldots$, 
$\{p_1,p_2,\ldots,p_{m-1}\}\in A_1$; $p_m\in A_2$; $\{q_1,q_2,\ldots,$ $q_{m-1}\}\in A_1$; $q_m\in A_3\cup A_4\cup\cdots\cup A_k$,$\ldots$, 
$\{p_1,p_2,\ldots,p_{m-1}\}\in A_k$; $p_m\in A_1$; $\{q_1,q_2,\ldots,q_{m-1}\}\in A_k$; $q_m\in A_2\cup A_3\cup\cdots\cup A_{k-1}$,$\ldots$, 
$\{p_1,p_2,\ldots,p_{m-2}\}\in A_1$; $p_{m-1}\in A_2$; $p_m\in A_3$; $\{q_1,q_2,\ldots,q_{m-2}\}\in A_1$; $q_{m-1}\in A_2$; $q_m\in A_4\cup A_5\cup\cdots\cup A_k$,
$\ldots$, $p_1\in A_1$; $p_2\in A_2$; $\ldots$; $p_{m-1}\in A_{k-2}$; $p_m\in A_k$; $q_1\in A_1$; $q_2\in A_2$; $\ldots$; $q_{m-1}\in A_{k-2}$; $q_m\in A_{k-1}$ 
is found to be 
\small
\begin{align}
&|\rho_{{2^{p_1}+2^{p_2}+\cdots+2^{p_m}+1},{2^{q_1}+2^{q_2}+\cdots+2^{q_m}+1}}| \notag\\
&=\sqrt{\rho_{{2^{p_1}+2^{p_2}+\cdots+2^{p_{m-1}}+1},{2^{p_1}+2^{p_2}+\cdots+2^{p_{m-1}}+1}} \rho_{{2^{q_1}+2^{q_2}+\cdots+2^{q_{m+1}}+1},{2^{q_1}+2^{q_2}+\cdots+2^{q_{m+1}}+1}}} \label{app3}
\end{align}
\normalsize
with the constraints $p_r, q_s \in \{ 0,1,\ldots,N-1\}$, $1\leq r\leq m-1$, $1\leq s\leq m+1$. 
$p_1 < p_2 < \cdots < p_{m-1}$, $q_1 < q_2 < \cdots < q_{m+1}$, 
$\{p_1,p_2,\ldots,p_{m-1}\} \cap \{q_1,q_2,\ldots,q_{m+1}\}$ $= \{p_1,p_2,\ldots,p_{m-1}\}$. 

Combining the above two cases with all possibilities provide   
\tiny
\begin{align}
&\sum_{p_1=m-1}^{N-1} \sum_{p_2=m-2}^{N-2} \ldots \sum_{p_{m-1}=1}^{N-(m-1)}\sum_{p_m=0}^{N-m} 
\sum_{q_1=m-1}^{N-1} \sum_{q_2=m-2}^{N-2}\ldots \sum_{q_{m-1}=1}^{N-(m-1)} \sum_{q_m=0}^{N-m} \notag \\
&\qquad\qquad|\rho_{{2^{p_1}+2^{p_2}+\ldots+2^{p_m}+1},{2^{q_1}+2^{q_2}+\ldots+2^{q_m}+1}}| \notag \\
&= \Bigg(\sum_{p_1=m-1,\atop p_1\in A_1}^{N-1} \sum_{p_2=m-2,\atop p_2\in A_1}^{N-2} \ldots \sum_{p_{m-1}=1,\atop p_{m-1}\in A_1}^{N-(m-1)} \sum_{p_m=0,\atop p_m\in A_1}^{N-m} \sum_{q_1=m-1,\atop q_1\in A_1}^{N-1} 
\sum_{q_2=m-2,\atop q_2\in A_1}^{N-2}\ldots \sum_{q_{m-1}=1,\atop q_{m-1}\in A_1}^{N-(m-1)} \sum_{q_m=0, \atop q_m \in A_2\cup A_3\cup \cdots \cup A_k}^{N-m} \notag \\
&\qquad |\rho_{{2^{p_1}+2^{p_2}+\ldots+2^{p_m}+1},{2^{q_1}+2^{q_2}+\ldots+2^{q_m}+1}}| \notag \\
&+ \sum_{p_1=m-1,\atop p_1\in A_2}^{N-1} \sum_{p_2=m-2,\atop p_2\in A_2}^{N-2} \ldots \sum_{p_{m-1}=1,\atop p_{m-1}\in A_2}^{N-(m-1)} \sum_{p_m=0,\atop p_m\in A_2}^{N-m} 
\sum_{q_1=m-1,\atop q_1\in A_2}^{N-1} \sum_{q_2=m-2,\atop q_2\in A_2}^{N-2}\ldots \sum_{q_{m-1}=1,\atop q_{m-1}\in A_2}^{N-(m-1)} \sum_{q_m=0,\atop q_m \in A_1\cup A_3\cup\cdots\cup A_k}^{N-m} \notag\\
&\qquad |\rho_{{2^{p_1}+2^{p_2}+\ldots+2^{p_m}+1},{2^{q_1}+2^{q_2}+\ldots+2^{q_m}+1}}| \notag \\ 
&+\cdots+ \sum_{p_1=m-1,\atop p_1\in A_k}^{N-1} \sum_{p_2=m-2,\atop p_2\in A_k}^{N-2} \ldots \sum_{p_{m-1}=1,\atop p_{m-1}\in A_k}^{N-(m-1)}
\sum_{p_m=0,\atop p_m\in A_k}^{N-m} 
\sum_{q_1=m-1,\atop q_1\in A_k}^{N-1} \sum_{q_2=m-2,\atop q_2\in A_k}^{N-2}\ldots \sum_{q_{m-1}=1,\atop q_{m-1}\in A_k}^{N-(m-1)} \sum_{q_m=0,\atop q_m \in A_1\cup A_2\cup \cdots\cup A_{k-1}}^{N-m} \notag\\
&\qquad |\rho_{{2^{p_1}+2^{p_2}+\ldots+2^{p_m}+1},{2^{q_1}+2^{q_2}+\ldots+2^{q_m}+1}}| \notag \\
&+\sum_{p_1=m-1,\atop p_1\in A_1}^{N-1} \sum_{p_2=m-2,\atop p_2\in A_1}^{N-2} \ldots \sum_{p_{m-1}=1,\atop p_{m-1}\in A_1}^{N-(m-1)} 
\sum_{p_m=0, \atop p_m\in A_2}^{N-m}  
\sum_{q_1=m-1,\atop q_1\in A_1}^{N-1} \sum_{q_2=m-2,\atop q_2\in A_1}^{N-2}\ldots \sum_{q_{m-1}=1,\atop q_{m-1}\in A_1}^{N-(m-1)} \sum_{q_m=0 \atop q_m\in A_3\cup A_4\cup \cdots\cup A_k}^{N-m} \notag\\ 
&\qquad |\rho_{{2^{p_1}+2^{p_2}+\ldots+2^{p_m}+1},{2^{q_1}+2^{q_2}+\ldots+2^{q_m}+1}}| \notag  \\
&+\cdots+\sum_{p_1=m-1,\atop p_1\in A_1}^{N-1} \sum_{p_2=m-2,\atop p_2\in A_1}^{N-2} \ldots \sum_{p_{m-1}=1,\atop p_{m-1}\in A_1}^{N-(m-1)} 
\sum_{p_m=0, \atop p_m\in A_k}^{N-m}  
\sum_{q_1=m-1,\atop q_1\in A_1}^{N-1} \sum_{q_2=m-2,\atop q_2\in A_1}^{N-2}\ldots \sum_{q_{m-1}=1,\atop q_{m-1}\in A_1}^{N-(m-1)} \sum_{q_m=0 \atop q_m\in A_2\cup A_3\cup \cdots\cup A_{k-1}}^{N-m} \notag\\
&\qquad |\rho_{{2^{p_1}+2^{p_2}+\ldots+2^{p_m}+1},{2^{q_1}+2^{q_2}+\ldots+2^{q_m}+1}}| \notag  \\
&+\cdots+\sum_{p_1=m-1,\atop p_1\in A_k}^{N-1} \sum_{p_2=m-2,\atop p_2\in A_k}^{N-2} \ldots \sum_{p_{m-1}=1,\atop p_{m-1}\in A_k}^{N-(m-1)} 
\sum_{p_m=0, \atop p_m\in A_1}^{N-m}  
\sum_{q_1=m-1,\atop q_1\in A_k}^{N-1} \sum_{q_2=m-2,\atop q_2\in A_k}^{N-2}\ldots \sum_{q_{m-1}=1,\atop q_{m-1}\in A_k}^{N-(m-1)} \sum_{q_m=0 \atop q_m\in A_2\cup A_3\cup \cdots\cup A_{k-1}}^{N-m} \notag\\
&\qquad |\rho_{{2^{p_1}+2^{p_2}+\ldots+2^{p_m}+1},{2^{q_1}+2^{q_2}+\ldots+2^{q_m}+1}}| \notag  \\
&+\cdots+\sum_{p_1=m-1,\atop p_1\in A_1}^{N-1}\ldots \sum_{p_{m-2}=2,\atop p_{m-2}\in A_1}^{N-(m-2)}  \sum_{p_{m-1}=1,\atop p_{m-1}\in A_2}^{N-(m-1)} 
\sum_{p_m=0, \atop p_m\in A_3}^{N-m}  
\sum_{q_1=m-1,\atop q_1\in A_1}^{N-1}\ldots \sum_{q_{m-2}=2,\atop q_{m-2}\in A_1}^{N-(m-2)}\sum_{q_{m-1}=1,\atop q_{m-1}\in A_2}^{N-(m-1)} \sum_{q_m=0 \atop q_m\in A_4\cup A_5\cup \cdots\cup A_k}^{N-m} \notag\\
&\qquad |\rho_{{2^{p_1}+2^{p_2}+\ldots+2^{p_m}+1},{2^{q_1}+2^{q_2}+\ldots+2^{q_m}+1}}| \notag  \\ 
&+\cdots+\sum_{p_1=m-1,\atop p_1\in A_1}^{N-1}\sum_{p_2=m-2,\atop p_2\in A_2}^{N-2} \ldots \sum_{p_{m-1}=1,\atop p_{m-1}\in A_{k-2}}^{N-(m-1)} 
\sum_{p_m=0, \atop p_m\in A_k}^{N-m}  
\sum_{q_1=m-1,\atop q_1\in A_1}^{N-1}\sum_{q_2=m-2,\atop q_2\in A_2}^{N-2}\ldots\sum_{q_{m-1}=1,\atop q_{m-1}\in A_{k-2}}^{N-(m-1)} \sum_{q_m=0 \atop q_m\in A_{k-1}}^{N-m} 
\notag\\ & \qquad |\rho_{{2^{p_1}+2^{p_2}+\ldots+2^{p_m}+1},{2^{q_1}+2^{q_2}+\ldots+2^{q_m}+1}}| \Bigg) \notag  \\
&+\cdots+\Bigg(\sum_{p_1=m-1,\atop p_1\in A_1}^{N-1}\sum_{p_2=m-2,\atop p_2\in A_2}^{N-2} \ldots \sum_{p_{m-1}=1,\atop p_{m-1}\in A_{k-1}}^{N-(m-1)} 
\sum_{p_m=0, \atop p_m\in A_k}^{N-m}  
\sum_{q_1=m-1,\atop q_1\in A_1}^{N-1}\sum_{q_2=m-2,\atop q_2\in A_2}^{N-2}\ldots\sum_{q_{m-1}=1,\atop q_{m-1}\in A_{k-1}}^{N-(m-1)} \sum_{q_m=0 \atop q_m\in A_k}^{N-m}
\notag\\ &\qquad |\rho_{{2^{p_1}+2^{p_2}+\ldots+2^{p_m}+1},{2^{q_1}+2^{q_2}+\ldots+2^{q_m}+1}}| \notag  \\
&+\cdots+\sum_{p_1=m-1,\atop p_1\in A_1}^{N-1}\sum_{p_2=m-2,\atop p_2\in A_1}^{N-2} \ldots \sum_{p_{m-1}=1,\atop p_{m-1}\in A_k}^{N-(m-1)} 
\sum_{p_m=0, \atop p_m\in A_k}^{N-m}  
\sum_{q_1=m-1,\atop q_1\in A_1}^{N-1}\sum_{q_2=m-2,\atop q_2\in A_1}^{N-2}\ldots\sum_{q_{m-1}=1,\atop q_{m-1}\in A_k}^{N-(m-1)} \sum_{q_m=0 \atop q_m\in A_k}^{N-m} 
\notag\\
& \qquad |\rho_{{2^{p_1}+2^{p_2}+\ldots+2^{p_m}+1},{2^{q_1}+2^{q_2}+\ldots+2^{q_m}+1}}| \notag \\
&+\cdots+\sum_{p_1=m-1,\atop p_1\in A_k}^{N-1}\sum_{p_2=m-2,\atop p_2\in A_k}^{N-2} \ldots \sum_{p_{m-1}=1,\atop p_{m-1}\in A_k}^{N-(m-1)} 
\sum_{p_m=0, \atop p_m\in A_k}^{N-m}  
\sum_{q_1=m-1,\atop q_1\in A_k}^{N-1}\sum_{q_2=m-2,\atop q_2\in A_k}^{N-2}\ldots\sum_{q_{m-1}=1,\atop q_{m-1}\in A_k}^{N-(m-1)} \sum_{q_m=0 \atop q_m\in A_k}^{N-m} \notag\\
&\qquad |\rho_{{2^{p_1}+2^{p_2}+\ldots+2^{p_m}+1},{2^{q_1}+2^{q_2}+\ldots+2^{q_m}+1}}| \Bigg) . 
\label{app4}
\end{align}
\normalsize

Substituting (\ref{app2}) and (\ref{app3}) in (\ref{app4}), we get 
\tiny
\begin{align}
&\sum_{p_1=m-1}^{N-1} \sum_{p_2=m-2}^{N-2} \ldots \sum_{p_{m-1}=1}^{N-(m-1)}\sum_{p_m=0}^{N-m} 
\sum_{q_1=m-1}^{N-1} \sum_{q_2=m-2}^{N-2}\ldots \sum_{q_{m-1}=1}^{N-(m-1)} \sum_{q_m=0}^{N-m} \notag\\
&\qquad\qquad|\rho_{{2^{p_1}+2^{p_2}+\ldots+2^{p_m}+1},{2^{q_1}+2^{q_2}+\ldots+2^{q_m}+1}}| \notag\\ 
&\leq \Bigg(\sum_{p_1=0,\atop p_1\in A_1}^{N-(m-1)} \sum_{p_2=1,\atop p_2\in A_1}^{N-(m-2)} \ldots \sum_{p_{m-1}=m-2,\atop p_{m-1}\in A_1}^{N-1} 
\sum_{q_1=0,\atop q_1\in A_1}^{N-(m+1)} \sum_{q_2=1,\atop q_2\in A_1}^{N-m}\ldots \sum_{q_m=m-1, \atop q_m \in A_1}^{N-2} \sum_{q_{m+1}=m,\atop q_{m+1}\in A_2\cup A_3\cup\cdots\cup A_k}^{N-1} \notag\\ 
&\sqrt{\rho_{{2^{p_1}+2^{p_2}+\cdots+2^{p_{m-1}}+1},{2^{p_1}+2^{p_2}+\cdots+2^{p_{m-1}}+1}} \rho_{{2^{q_1}+2^{q_2}+\cdots+2^{q_{m+1}}+1},{2^{q_1}+2^{q_2}+\cdots+2^{q_{m+1}}+1}}} \notag\\ 
&+\sum_{p_1=0,\atop p_1\in A_2}^{N-(m-1)} \sum_{p_2=1,\atop p_2\in A_2}^{N-(m-2)} \ldots \sum_{p_{m-1}=m-2,\atop p_{m-1}\in A_2}^{N-1} 
\sum_{q_1=0,\atop q_1\in A_2}^{N-(m+1)} \sum_{q_2=1,\atop q_2\in A_2}^{N-m}\ldots \sum_{q_m=m-1, \atop q_m \in A_2}^{N-2} \sum_{q_{m+1}=m,\atop q_{m+1}\in A_1\cup A_3\cup\cdots\cup A_k}^{N-1}  \notag\\ 
& \sqrt{\rho_{{2^{p_1}+2^{p_2}+\cdots+2^{p_{m-1}}+1},{2^{p_1}+2^{p_2}+\cdots+2^{p_{m-1}}+1}} \rho_{{2^{q_1}+2^{q_2}+\cdots+2^{q_{m+1}}+1},{2^{q_1}+2^{q_2}+\cdots+2^{q_{m+1}}+1}}} \notag\\ 
&+\cdots+\sum_{p_1=0,\atop p_1\in A_k}^{N-(m-1)} \sum_{p_2=1,\atop p_2\in A_k}^{N-(m-2)} \ldots \sum_{p_{m-1}=m-2,\atop p_{m-1}\in A_k}^{N-1} 
\sum_{q_1=0,\atop q_1\in A_k}^{N-(m+1)} \sum_{q_2=1,\atop q_2\in A_k}^{N-m}\ldots \sum_{q_m=m-1, \atop q_m \in A_k}^{N-2} \sum_{q_{m+1}=m,\atop q_{m+1}\in A_1\cup A_2\cup\cdots\cup A_{k-1}}^{N-1}  \notag\\ 
& \sqrt{\rho_{{2^{p_1}+2^{p_2}+\cdots+2^{p_{m-1}}+1},{2^{p_1}+2^{p_2}+\cdots+2^{p_{m-1}}+1}} \rho_{{2^{q_1}+2^{q_2}+\cdots+2^{q_{m+1}}+1},{2^{q_1}+2^{q_2}+\cdots+2^{q_{m+1}}+1}}} \notag\\ 
&+\cdots+\sum_{p_1=0,\atop p_1\in A_k}^{N-(m-1)} \sum_{p_2=1,\atop p_2\in A_k}^{N-(m-2)} \ldots \sum_{p_{m-1}=m-2,\atop p_{m-1}\in A_k}^{N-1} 
\sum_{q_1=0,\atop q_1\in A_k}^{N-(m+1)} \ldots \sum_{q_{m-1}=m-2, \atop q_{m-1} \in A_k}^{N-3} \sum_{q_m=m-1, \atop q_m \in A_1\cup \cdots\cup A_{k-1}}^{N-2} \sum_{q_{m+1}=m,\atop q_{m+1}\in A_1\cup \cdots\cup A_{k-1}}^{N-1}  \notag\\ 
& \sqrt{\rho_{{2^{p_1}+2^{p_2}+\cdots+2^{p_{m-1}}+1},{2^{p_1}+2^{p_2}+\cdots+2^{p_{m-1}}+1}} \rho_{{2^{q_1}+2^{q_2}+\cdots+2^{q_{m+1}}+1},{2^{q_1}+2^{q_2}+\cdots+2^{q_{m+1}}+1}}} \notag\\ 
&+\cdots+\sum_{p_1=0,\atop p_1\in A_1}^{N-(m-1)} \ldots \sum_{p_{m-2}=m-3,\atop p_{m-2}\in A_1}^{N-2}\sum_{p_{m-1}=m-2,\atop p_{m-1}\in A_2}^{N-1} 
\sum_{q_1=0,\atop q_1\in A_1}^{N-(m+1)} \ldots \sum_{q_{m-1}=m-2, \atop q_{m-1} \in A_2}^{N-3} \sum_{q_m=m-1, \atop q_m \in 
A_3\cup \cdots\cup A_k}^{N-2} \sum_{q_{m+1}=m,\atop q_{m+1}\in A_3\cup \cdots\cup A_k}^{N-1} \notag\\ 
&\sqrt{\rho_{{2^{p_1}+2^{p_2}+\cdots+2^{p_{m-1}}+1},{2^{p_1}+2^{p_2}+\cdots+2^{p_{m-1}}+1}} \rho_{{2^{q_1}+2^{q_2}+\cdots+2^{q_{m+1}}+1},{2^{q_1}+2^{q_2}+\cdots+2^{q_{m+1}}+1}}} \notag\\ 
&+\cdots+\sum_{p_1=0,\atop p_1\in A_k}^{N-(m-1)} \ldots \sum_{p_{m-2}=m-3,\atop p_{m-2}\in A_k}^{N-2}\sum_{p_{m-1}=m-2,\atop p_{m-1}\in A_{k-1}}^{N-1} 
\sum_{q_1=0,\atop q_1\in A_k}^{N-(m+1)} \ldots \sum_{q_{m-1}=m-2, \atop q_{m-1} \in A_{k-1}}^{N-3} \sum_{q_m=m-1, \atop q_m \in A_1\cup \cdots\cup A_{k-2}}^{N-2} \sum_{q_{m+1}=m,\atop q_{m+1}\in A_1\cup \cdots\cup A_{k-2}}^{N-1} \notag\\
&\sqrt{\rho_{{2^{p_1}+2^{p_2}+\cdots+2^{p_{m-1}}+1},{2^{p_1}+2^{p_2}+\cdots+2^{p_{m-1}}+1}} \rho_{{2^{q_1}+2^{q_2}+\cdots+2^{q_{m+1}}+1},{2^{q_1}+2^{q_2}+\cdots+2^{q_{m+1}}+1}}} \notag\\ 
&+\cdots+\sum_{p_1=0,\atop p_1\in A_1}^{N-(m-1)} \sum_{p_2=1,\atop p_{2}\in A_2}^{N-(m-2)} \ldots \sum_{p_{m-1}=m-2,\atop p_{m-1}\in A_{k-2}}^{N-1} 
\sum_{q_1=0,\atop q_1\in A_1}^{N-(m+1)} \sum_{q_2=1, \atop q_2 \in A_2}^{N-m} \ldots \sum_{q_{m-1}=m-2, \atop q_{m-1} \in A_{k-2}}^{N-3} \sum_{q_m=m-1, \atop q_m \in A_{k-1}}^{N-2} \sum_{q_{m+1}=m,\atop q_{m+1}\in A_1\cup \cdots\cup A_{k-3}}^{N-1}  \notag\\ 
& \sqrt{\rho_{{2^{p_1}+2^{p_2}+\cdots+2^{p_{m-1}}+1},{2^{p_1}+2^{p_2}+\cdots+2^{p_{m-1}}+1}} \rho_{{2^{q_1}+2^{q_2}+\cdots+2^{q_{m+1}}+1},{2^{q_1}+2^{q_2}+\cdots+2^{q_{m+1}}+1}}} \Bigg) \notag\\ 
&+\cdots+\Bigg(\sum_{p_1=0,\atop p_1\in A_1}^{N-m}\sum_{p_2=1,\atop p_2\in A_2}^{N-(m-1)} \ldots \sum_{p_{m-1}=m-2,\atop p_{m-1}\in A_{k-1}}^{N-2} 
\sum_{p_m=m-1, \atop p_m\in A_k}^{N-1}  
\sum_{q_1=0,\atop q_1\in A_1}^{N-m}\sum_{q_2=1,\atop q_2\in A_2}^{N-(m-1)}\ldots\sum_{q_{m-1}=m-2,\atop q_{m-1}\in A_{k-1}}^{N-2} \sum_{q_m=m-1 \atop q_m\in A_k}^{N-1} 
\notag\\ 
& \frac{\rho_{{2^{p_1}+2^{p_2}+\cdots+2^{p_m}+1},{2^{p_1}+2^{p_2}+\cdots+2^{p_m}+1}}+\rho_{{2^{q_1}+2^{q_2}+\cdots+2^{q_m}+1},{2^{q_1}+2^{q_2}+\cdots+2^{q_m}+1}}}{2} 
\notag \\
&+\cdots+\sum_{p_1=0,\atop p_1\in A_1}^{N-m}\sum_{p_2=1,\atop p_2\in A_1}^{N-(m-1)} \ldots \sum_{p_{m-1}=m-2,\atop p_{m-1}\in A_k}^{N-2} 
\sum_{p_m=m-1, \atop p_m\in A_k}^{N-1}  
\sum_{q_1=0,\atop q_1\in A_1}^{N-m}\sum_{q_2=1,\atop q_2\in A_1}^{N-(m-1)}\ldots\sum_{q_{m-1}=m-2,\atop q_{m-1}\in A_k}^{N-2} \sum_{q_m=m-1 \atop q_m\in A_k}^{N-1} 
\notag\\ 
&  \frac{\rho_{{2^{p_1}+2^{p_2}+\cdots+2^{p_m}+1},{2^{p_1}+2^{p_2}+\cdots+2^{p_m}+1}}+\rho_{{2^{q_1}+2^{q_2}+\cdots+2^{q_m}+1},{2^{q_1}+2^{q_2}+\cdots+2^{q_m}+1}}}{2} 
\notag \\
&+\cdots+\sum_{p_1=0,\atop p_1\in A_k}^{N-m}\sum_{p_2=1,\atop p_2\in A_k}^{N-(m-1)} \ldots \sum_{p_{m-1}=m-2,\atop p_{m-1}\in A_k}^{N-2} 
\sum_{p_m=m-1, \atop p_m\in A_k}^{N-1}  
\sum_{q_1=0,\atop q_1\in A_k}^{N-m}\sum_{q_2=1,\atop q_2\in A_k}^{N-(m-1)}\ldots\sum_{q_{m-1}=m-2,\atop q_{m-1}\in A_k}^{N-2} \sum_{q_m=m-1 \atop q_m\in A_k}^{N-1} 
\notag\\ 
& \frac{\rho_{{2^{p_1}+2^{p_2}+\cdots+2^{p_m}+1},{2^{p_1}+2^{p_2}+\cdots+2^{p_m}+1}}+\rho_{{2^{q_1}+2^{q_2}+\cdots+2^{q_m}+1},{2^{q_1}+2^{q_2}+\cdots+2^{q_m}+1}}}{2}\Bigg).
\label{app5}
\end{align}
\normalsize
In the left hand side we assign the following constraints, that is, 
$p_k, q_k \in \{ 0,1,\ldots,N-1\}$, $1\leq k\leq m$, 
$p_m < p_{m-1} < \cdots < p_2 < p_1$, $q_m < q_{m-1} < \cdots < q_2 < q_1$, 
$q_m\geq p_m, q_{m-1}\geq p_{m-1},\ldots,q_1\geq p_1$, $(p_1,p_2,\ldots,p_{m-1},p_m)\neq (q_1,q_2,\ldots,q_{m-1},q_m)$.  
The set $\{p_1,p_2,\ldots,p_m\} \cap \{q_1,q_2,\ldots,q_m\}$ has exactly $(m-1)$ elements. 
In the right hand side, the terms within in the first parenthesis possess the following constraints, that is, 
$p_r, q_s \in \{ 0,1,\ldots,N-1\}$, $1\leq r\leq m-1$, $1\leq s\leq m+1$. 
$p_1 < p_2 < \cdots < p_{m-1}$, $q_1 < q_2 < \cdots < q_{m+1}$, 
$\{p_1,p_2,\ldots,p_{m-1}\} \cap \{q_1,q_2,\ldots,q_{m+1}\}$ $= \{p_1,p_2,\ldots,p_{m-1}\}$. 
The terms within the second parenthesis have the following constraints, that is, 
$p_1 < p_2 < \cdots < p_m$, $q_1 < q_2 < \cdots < q_m$.  

We simplify the above expression and introduce new labelling in order to distinguish 
the terms that appear in the left hand side from right hand side. 
The final expression (\ref{app5}) read now
\renewcommand{\theequation}{\arabic{equation}}
\begin{align}
\setcounter{equation}{4}
&\displaystyle\sum_{p_1=m-1}^{N-1} \displaystyle\sum_{p_2=m-2}^{N-2} \ldots \displaystyle\sum_{p_{m-1}=1}^{N-(m-1)}\displaystyle\sum_{p_m=0}^{N-m} 
\displaystyle\sum_{q_1=m-1}^{N-1} \displaystyle\sum_{q_2=m-2}^{N-2} 
\ldots \displaystyle\sum_{q_{m-1}=1}^{N-(m-1)}\displaystyle\sum_{q_m=0}^{N-m} \notag\\ 
&\qquad\qquad\qquad |\rho_{{2^{p_1}+2^{p_2}+\ldots+2^{p_m}+1},{2^{q_1}+2^{q_2}+\ldots+2^{q_m}+1}}|\notag\\
&\leq \displaystyle\sum_{i_1=0}^{N-(m-1)} \displaystyle\sum_{i_2=1}^{N-(m-2)} \ldots \displaystyle\sum_{i_{m-1}=m-2}^{N-1} \displaystyle\sum_{j_1=0}^{N-(m+1)} 
\displaystyle\sum_{j_2=1}^{N-m} \ldots \displaystyle\sum_{j_{m+1}=m}^{N-1}\notag\\ 
&\sqrt{\rho_{{2^{i_1}+2^{i_2}+\ldots+2^{i_{m-1}}+1},{2^{i_1}+2^{i_2}+\ldots+2^{i_{m-1}}+1}} 
\rho_{{2^{j_1}+2^{j_2}+\ldots+2^{j_{m+1}}+1},{2^{j_1}+2^{j_2}+\ldots+2^{j_{m+1}}+1}}} \notag \\
& +\left(\frac{N-k}{2}\right)\displaystyle\sum_{r_1=0}^{N-m} \displaystyle\sum_{r_2=1}^{N-(m-1)}\ldots \displaystyle\sum_{r_{m-1}=m-2}^{N-2} 
\displaystyle\sum_{r_{m}=m-1}^{N-1} \notag\\
&\qquad\qquad\qquad\qquad \rho_{{2^{r_1}+2^{r_2}+\ldots+2^{r_m}+1},{2^{r_1}+2^{r_2}+\ldots+2^{r_m}+1}}, 
\end{align}
with the parameter contraints mentioned in section \ref{sec3}.   
The inequality (\ref{dik}) holds for $k$-separable $N$-qubit pure state with $m$ excitations. 
By using Cauchy inequality and carrying out simple algebras, we can also show that 
(\ref{dik}) holds for $k$-separable $N$-qubit mixed states \cite{gao2011}. 
The expression (\ref{dik}) is the required condition to identify the non-$k$-separability of 
$N$-qubit states. 



\begin{thebibliography}{}

\bibitem{horo2009}
Horodecki, R., Horodecki, P., Horodecki, M., Horodecki, K.: Rev. Mod. Phys. {\bf 81}, 865 (2009) 
\bibitem{guhne2009}
G\"uhne, O., T\'oth, G.: Phys. Rep. {\bf 474}, 1 (2009)

\bibitem{guhne2005}
G\"uhne, O., T\'oth, G., Briegel, H.J.: New J. Phys. {\bf 7}, 229 (2005)  

\bibitem{novo2013} 
Novo, L., Moroder, T., G\"uhne, O.: Phys. Rev. A {\bf 88}, 012305 (2013)  

\bibitem{dick1954}
Dicke, R.H.: Phys. Rev. {\bf 93}, 99 (1954)  

\bibitem{toth2007}
T\'oth, G.: J. Opt. Soc. Am. B {\bf 24}, 275 (2007) 

\bibitem{berg2013}
Bergmann, M., G\"uhne, O.: J. Phys. A {\bf 46}, 385304 (2013)

\bibitem{haff2005}
H\"affner, H., et al.: Nature (London) {\bf 438}, 643 (2005) 

\bibitem{kies2007}
Kiesel, N., Schmid, C., T\'oth, G., Solano, E., Weinfurter, H.: Phys. Rev. Lett. {\bf 98}, 063604 (2007)

\bibitem{prev2009}
Prevedel, R., Cronenberg, G., Tame, M.S., Paternostro, M., Walther, P., Kim, M.S., Zeilinger, A.: Phys. Rev. Lett. {\bf 103}, 020503 (2009)  

\bibitem{wiec2009}
Wieczorek, W., Krischek, R., Kiesel, N., Michelberger, P., T\'oth, G., Weinfurter, H.: Phys. Rev. Lett. {\bf 103}, 020504 (2009) 

\bibitem{lini2008}
Linington, I.E., Vitanov, N.V.: Phys. Rev. A {\bf 77}, 010302(R) (2008) 

\bibitem{toyo2011}
Toyoda, K., Watanabe, T., Kimura, T., Nomura, S., Haze, S., Urabe, S.: Phys. Rev. A {\bf 83}, 022315 (2011)   

\bibitem{guhn2008}
G\"uhne, O., Bodoky, F., Blaauboer, M.: Phys. Rev. A {\bf 78}, 060301(R) (2008)   

\bibitem{mura1999}
Murao, M., Jonathan, D., Plenio, M.B., Vedral, V.: Phys. Rev. A \textbf{59}, 156 (1999)  

\bibitem{hill1999}
Hillery, M., Bu\v{z}ek, V., Berthiaume, A.: Phys. Rev. A  \textbf{59}, 1829 (1999) 

\bibitem{hube2011}
Huber, M., Erker, P., Schimpf, H., Gabriel, A., Hiesmayr, B.: Phys. Rev. A \textbf{83}, 040301(R) (2011) 

\bibitem{duan2011}
Duan, L-M.:  Phys. Rev. Lett. {\bf 107}, 180502 (2011)   

\bibitem{luck2014}
L\"ucke, B., Peise, J., Vitagliano, G., Arlt, J., Santos, L., T\'oth, G., Klempt, C.: Phys. Rev. Lett. {\bf 112}, 155304 (2014)  

\bibitem{gao2014}
Gao, T., Yan, F., van Enk, S.J.: Phys. Rev. Lett. {\bf 112}, 180501 (2014)  

\bibitem{anan2015} 
Ananth, N., Chandrasekar, V.K., Senthilvelan, M.: Eur. Phys. J. D \textbf{69}, 56 (2015) 

\bibitem{gabr2010}
Gabriel, A., Hiesmeyr, B.C., Huber, M.: Quantum Inf. Comput. \textbf{10}, 0829 (2010)  

\bibitem{agra2006}
Agrawal, P., Pati, A.: Phys. Rev. A \textbf{74}, 062320 (2006)

\bibitem{zheng2006}
Zheng, S-B.: Phys. Rev. A {\bf 74}, 054303 (2006)

\bibitem{hond2006}
D'Hondt, E., Panangaden, P.: Quantum Inf. Comput. {\bf 6}, 173 (2006) 

\bibitem{gao2013}
Gao, T., Hong, Y., Lu, Y., Yan, F.: Europhys. Lett. \textbf{104}, 20007 (2013) 

\bibitem{guhne2010}
G\"uhne, O., Seevinck, M.: New J. Phys. \textbf{12}, 053002 (2010) 

\bibitem{gao2011}
Gao, T., Hong, Y.: Eur. Phys. J. D \textbf{61}, 765 (2011)  

\bibitem{hube2010}
Huber, M., Mintert, F., Gabriel, A., Hiesmayr, B.C.:  Phys. Rev. Lett. \textbf{104}, 210501 (2010)

\bibitem{kim2008}
Kim, J.S., Sanders, B.C.: J. Phys. A \textbf{41}, 495301 (2008)

\bibitem{seev2008}
Seevinck, M., Uffink, J.: Phys. Rev. A \textbf{78}, 032101 (2008)   

\bibitem{chiu2010}
Chiurib, A., Vallone, G., Bruno, N., Macchiavello, C., Bru\ss, ~D., Mataloni, P.: Phys. Rev. Lett. {\bf 105}, 250501 (2010)  

\bibitem{guhne2003}
G\"uhne, O., Hyllus, P.: Int. J. Theor. Phys. {\bf 42}, 1001 (2003)  

\bibitem{guhn2007}
G\"uhne, O., Lu, C-Y., Gao, W-B., Pan, J-W.: Phys. Rev. A \textbf{76}, 030305(R) (2007)  

\bibitem{gao2010}
Gao, T., Hong, Y.: Phys. Rev. A \textbf{82}, 062113 (2010)  

\bibitem{bert2007}
Bertlmann, R.A.,  Krammer, P.: arXiv:quant-ph/07061743

\bibitem{zhao2014}
Zhao, H., Fei, S-M., Fan, J., Wang, Z-X.: Int. J. Quan. Inf. \textbf{12}, 1450013 (2014) 



\end{thebibliography}


\end{document}